# Comment: Expert Elicitation for Reliable System Design

**Norman Fenton and Martin Neil**

The paper "Expert Elicitation for Reliable System Design" by Bedford, Quigley and Walls is timely and significant for three reasons:

1. It addresses the importance of expert elicitation in systems design and the statistical and practical challenges faced when trying to use expert judgements in a way that is consistent with established approaches based on statistical reliability testing.
2. It rightly focuses our attention on the need for a holistic approach to reliability evaluation that goes beyond analysis of single projects to also include information from "softer" sources such as design and operational use.
3. It recognizes the emerging importance of Bayesian methods in providing the "uncertainty calculus" to combine evidence from experts with statistical reliability data in such a way that system reliability assessments and forecasts can grow and evolve as a system changes throughout its life.

Our own research and experience support many of the key thrusts of the authors' ideas. For the last ten years we have been applying Bayesian methods—more specifically, Bayesian networks (which the authors refer to in Section 4.2.3)—to a wide variety of problem areas (see, e.g., Neil, Malcolm and Shaw, 2003, and Fenton et al., 2004). This includes system dependability evaluation, of which the best known example is the Transport Reliability Assessment Calculation System (TRACS) (Neil, Fenton, Forey and Harris, 2001); this is an early exemplar of the meta modeling frameworks cited by the authors in Section 4.1. We have found Bayesian methods to be most beneficial to the types of problems mentioned by the authors, including the issue of making trade-offs between reliability and other system objectives like functionality and cost (something we examined in detail for software systems in Fenton et al., 2004).

We have a number of additional observations to make about the paper:

Very often reliability assessments are carried out by a client (rather than the design authority) or by a procurement agency on behalf of the client. In this case, the expert is not the designer but a customer, and the impact of this is more general than the authors appear to suggest in Table 1. Such customers may have relevant operational reliability experience gained from use of similar products from this or different suppliers and will, quite correctly, want to use this experience to best effect either to reduce testing effort or to select suppliers at the procurement stage. Other situations spring to mind where a different perspective would give rise to additional problems and challenges, such as COTS (commercial off the shelf systems).

There can be a paucity of empirical data for mission and safety critical systems simply because the systems may be novel or the top events may be rare. Probabilistic risk assessment methods aside, this problem often forces practitioners to borrow or adopt data from different sources, some of uncertain provenance, to help make a reliability claim based on some structured (or often unstructured) argument. Where data do exist, they may only be partially relevant for a number of reasons. For example, the data may be sourced from heterogeneous systems or may have been collected under different or uncontrolled conditions. Detailed statistical modeling is practically and economically infeasible in such "messy" situations, but nevertheless judgements have to be made. In practice these decisions can be a black art, involving opaque assumptions and unchecked subjectivity, but in our experience Bayesian methods can help bring some rigor and structure. More importantly, they also encourage transparency and allow uncertainties and assumptions to be modeled explicitly.

*Norman Fenton is Professor of Computer Science and Head of Risk Assessment and Decision Analysis Research, Computer Science Department, Queen Mary College, University of London, London E1 4NS, United Kingdom e-mail: norman@dcs.qmul.ac.uk. Martin Neil is Reader in Systems Risk at Queen Mary College, University of London and CTO of Agenda Ltd e-mail: martin@dcs.qmul.ac.uk.*

In TRACS (Neil, Fenton, Forey and Harris, 2001) we built a system that partially or wholly addresses some of the authors' aims with some success. Indeed the system remains in routine use by QinetiQ to assess the reliability of military vehicles throughout procurement, design, test and operational use. One of the original key motivations for TRACS was exactly the problem identified in Section 4.1 that traditional approaches to reliability prediction tend to be overly optimistic because they fail to take into account design and process factors. The TRACS architecture allows estimation of failure rates from families of components using a Bayesian hierarchical model and then aggregates these into a system level reliability distribution, which can then be updated, using Bayes' rule and likelihood data gathered at prototype test, system trial and preproduction stages. Crucially, at each stage a number of expert-based assessments are made to adjust the failure rate predictions based on qualitative estimates of design and manufacturing factors, including subcontractor competence, risk analysis quality, design documentation quality, staff reputation and skills. A hybrid Bayesian network is then used to fuse all of the information to provide a family of estimates and predictions throughout system life. The state of the art has moved on considerably since TRACS, and the Bayesian algorithms used in TRACS are now available commercially (AgenaRisk, 2006). As a result, model construction is now considerably faster and easier than it was when TRACS was first implemented in 1999.

The issue of expert elicitation is becoming increasingly relevant to extend and supplement six sigma approaches. For example, we have recently been working with Motorola to help complement their six sigma program by using Bayesian methods to represent expert judgements about the impact of fundamental organizational and process factors on downstream product reliability. This is commercially important because reliability problems often occur as a result of sources of systematic design variability, often itself caused by the ineffective management of outsourced suppliers and problems with communicating and implementing system requirements. These are issues that are not easily addressed by statistical process control techniques, nor are such techniques designed to address them, despite their importance. Based on this experience, a number of interesting research issues relevant to the paper spring to mind:

- Cultural conflict; that is, how do we persuade engineering experts to express Bayesian priors when the dominant culture of statistical process control is almost entirely data driven [which can lead to what Chapman calls a syndrome of objective irrationality (Chapman and Ward, 2000)]?
- What universal organizational and process drivers affect what industries and in what way?
- Can we assess the effects of process factors in quantitative terms or encourage the adoption of methodical collection and sharing of the necessary data?

The authors implicitly assume that the benefits of probability elicitation will only accrue in situations where there is already a highly developed reliability methodology to which new techniques can be added. In these situations there are already structure, methods and data, but what of those who need to assess reliability of products sourced from less mature organizations or where data collection by empirical means is economically infeasible? Here elicitation could, perhaps controversially, be used instead of traditional reliability methods. In this situation decisions would turn on "softer" issues, but would nevertheless be quantified and the prediction ultimately would be verifiable, at least in principle.

An additional key benefit of probability elicitation that was not covered in the paper is that it helps codify knowledge, making it available in the future for other projects or for other systems. This is important because reliability assessment is not just a one-off activity undertaken on a single system or project or even over the lifetime of such systems; it also addresses families of systems that change within a changing design organization or usage environment. From this perspective, elicitation should be seen as a knowledge management opportunity rather than as a technical problem to be solved in isolation. Such knowledge, if codified and trusted, could be reused at reduced cost on future projects and used to help communicate engineering judgement from engineering experts to novices.

The issue of bias in subjective probability elicitation (which the authors address in Section 3.2) has too often been used as an easy excuse not to do Bayesian modeling. We feel strongly that this issue has been overplayed—a good discussion of this can be found in Ayton and Pascoe (1996). Moreover, in our own work building Bayesian net models with domain experts, we have developed a range of techniques that minimize the effort required for probability elicitation. An example is the use of simple predefined distributions that cover most common situations that involve ordinal scale variables that are conditioned on other ordinal scale variables (Fenton and Neil, 2006).

Finally, we would like to congratulate the authors on writing such an interesting, wide ranging and thought provoking paper.

## REFERENCES

AGENARISK (2006). Available at www.agenarisk.com.
AYTON, P. and PASCOE, E. (1996). Bias in cognitive judgements? *Knowledge Engineering Review* **10** 21–41.
CHAPMAN, C. B. and WARD, S. C. (2000). Estimation and evaluation of uncertainty: A minimalist first pass approach. *International J. Project Management* **18** 369–383.
FENTON, N. E., MARSH, W., NEIL, M., CATES, P., FOREY, S. and TAILOR, M. (2004). Making resource decisions for software projects. In *Proc. 26th International Conference on Software Engineering 2004* 397–406. IEEE Computer Society, Washington.
FENTON, N. E. and NEIL, M. (2006). Using ranked nodes to model qualitative judgements in Bayesian networks. Available at www.dcs.qmw.ac.uk/~norman/papers/ranked_nodes%20v01.004.pdf.
NEIL, M., FENTON, N., FOREY, S. and HARRIS, R. (2001). Using Bayesian belief networks to predict the reliability of military vehicles. *IEE Computing and Control Engineering J.* **12** 11–20.
NEIL, M., MALCOLM, B. and SHAW, R. (2003). Modeling an air traffic control environment using Bayesian belief networks. Presented at 21st International System Safety Conference, Ottawa, ON, Canada.